\documentclass[iop]{emulateapj-rtx4}

\usepackage{epsfig}
\usepackage{psfrag}
\newcommand{\mypsfrag}[2]{\psfrag{#1}{{\footnotesize #2}}}
\usepackage{amsfonts}          
\usepackage{amssymb}           
\usepackage{amsmath}           
\usepackage{xspace}
\usepackage{apjfonts}
\usepackage{color}

\renewcommand{\Re}{\mathrm{Re}}
\renewcommand{\Im}{\mathrm{Im}}
\newcommand{\sgn}{\mathrm{sgn}}
\renewcommand{\vec}[1]{\ensuremath{{\bf #1}}}

\providecommand{\MD}{\textbf{MD}\xspace}
\providecommand{\FD}{\textbf{FD}\xspace}
\providecommand{\Mdip}{\ensuremath{{M}_\mathrm{dip}}\xspace}

\shorttitle{How far can minimal models explain the solar cycle?}
\shortauthors{Simitev and Busse}

\begin{document}


\title{How far can minimal models explain the solar cycle?}


\author{R. D. Simitev}
\affil{School of Mathematics and Statistics, University of Glasgow,
    Glasgow, G12 8QW, UK}
\and
\author{F. H. Busse}
\affil{Institute of Physics, University of Bayreuth, 95440 Bayreuth, Germany}


\begin{abstract}
A physically consistent model of magnetic field generation by
convection in a rotating spherical shell with a minimum of parameters
is applied to the Sun. Despite its unrealistic features the model
exhibits a number of properties resembling those observed on the
Sun. The model suggests that the large scale solar dynamo is dominated
by a non-axisymmetric $m=1$ component of the magnetic field. 
\end{abstract}





\section{\uppercase{Introduction}}

During the past two decades the understanding of the process of
magnetic field generation by motions of an electrically conducting
fluid has progressed rapidly and increasingly realistic models of
dynamos operating in planetary interiors have been achieved. Models of
the geodynamo, of the ancient Martian dynamo or of Mercury's dynamo
are primarily restricted by the lack of knowledge of certain
properties of the respective planetary core. The dynamo process
operating in the Sun, on the other hand, is still subject to
controversies with several competing proposals. The density variations
of several orders of magnitude in the solar convection zone and the
complications introduced by the compressibility of the fluid have
prevented so far a fully convincing model of the solar cycle. 

In this situation it can be helpful to consider the properties of a
{most simple} physically consistent model of  convection driven dynamos
in a rotating spherical shell in order to see which properties 
might resemble those of the solar cycle. Assuming a given radius ratio of
the spherical fluid shell the minimum number of dimensionless
parameters is four, namely the Rayleigh number $R$, the rotation
parameter $\tau$, the Prandtl number $P$ denoting the ratio of viscous
and thermal diffusivity, and the magnetic Prandtl number $P_m$
denoting the ratio of viscous and magnetic diffusivity.   

The success of numerical simulations of the generation of magnetic
fields by turbulent fluid motions relies of the concept of eddy
diffusivities which represent the diffusive actions of small scale
turbulence {that} remains unresolved in the numerical schemes. The
ratios of those eddy diffusivities are expressed by the values of $P$
and $P_m$ used in the numerical computations. Frequently the
assumption $P=P_m=1$ is made based on the argument that the turbulent
diffusion affects the transport of heat, momentum and magnetic flux in
the same way. This assumption is not correct, however, since the
transports of scalar and vector quantities are affected in different
ways \citep{Kay05,You03}. Moreover, the diffusivities in the absence of any
motion often differ by many orders of magnitude, - in particular when
the radiative transport is taken into account in the heat diffusivity
-, such that even highly turbulent fluids do not fully erase
differences in the effective diffusivities. 

Even if it is {assumed} that ratios of the effective diffusivities for
heat, momentum and magnetic flux do not differ much from unity in the
highly turbulent solar convection zone, it must be realized that
numerical simulations of convection driven dynamos depend sensitively
on the parameters $P$ and $P_m$ near their values of unity. In
particular it has been found that a transition between two different
types of dynamos occurs which is hysteretic as a function of $P$ and
$P_m$ \citep{sb09}. It thus appears to be of interest to
study simple convection driven dynamos in dependence on these two
parameters in addition to the Rayleigh number and the rotation
parameter. More realistic solar dynamo models involve many more
parameters and because of the high demand for numerical resolution the
opportunities for { comprehensive studies of parameter dependences}
are restricted.  

For the present study a numerical model for convection driven dynamos that
previously has been employed for investigations motivated by problems
of the geodynamo \cite[see, for example,][]{sb05} has been modified for solar applications. A much
thinner spherical shell is assumed and no-slip conditions at  the
inner boundary and stress-free conditions at the outer boundary are
used. The Boussinesq approximation has been retained even though it
is not appropriate for any realistic solar dynamo model and  causes
an unrealistic dependence of the solar differential rotation on
depth. Nevertheless, as we hope to demonstrate, there are properties
of the dynamo solutions that resemble those observed on the Sun. 
\begin{figure*}[t]
\mypsfrag{R}{$R$}
\mypsfrag{Pm}{$P_m$}
\mypsfrag{P}{$P$}
\mypsfrag{0.6}{\hspace{2mm}0.6}
\mypsfrag{0.8}{\hspace{2mm}0.8}
\mypsfrag{1}  {\hspace{2mm}1}
\mypsfrag{1.2}{\hspace{2mm}1.2}
\mypsfrag{1.4}{\hspace{2mm}1.4}
\mypsfrag{1.6}{\hspace{2mm}1.6}
\mypsfrag{1.8}{\hspace{2mm}1.8}
\mypsfrag{1.5}{\hspace{-4mm}1.5}
\mypsfrag{2}  {\hspace{-4mm}2}
\mypsfrag{3}  {\hspace{-4mm}3}
\mypsfrag{4}  {\hspace{-4mm}4}
\mypsfrag{5}  {\hspace{-4mm}5}
\mypsfrag{7}  {\hspace{-4mm}7}
\mypsfrag{10}  {\hspace{-4mm}10}
\mypsfrag{10e5}  {\hspace{-0mm}$10^5$}
\begin{center}
\epsfig{file=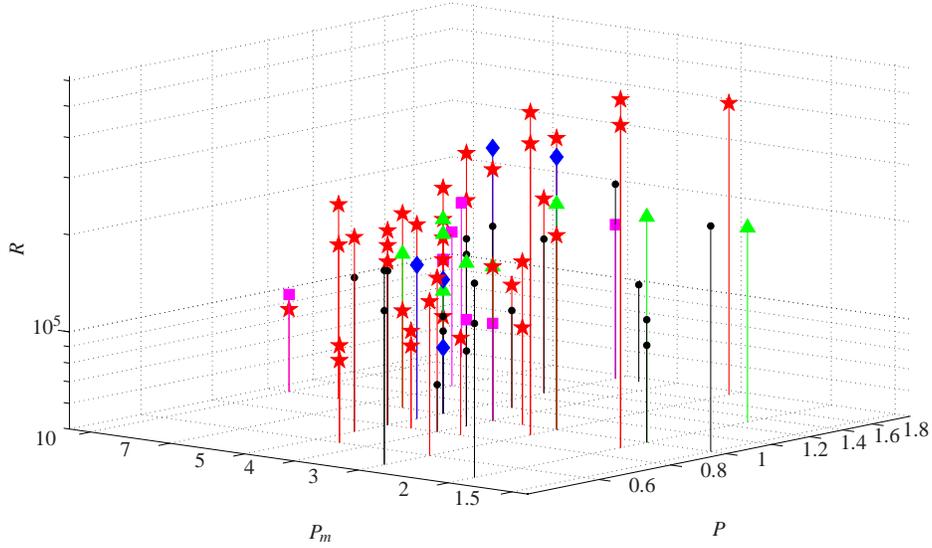,width=0.8\textwidth,clip=}
\end{center}
\caption[]{(Color online)
Convection-driven dynamos as a function of $R$, $P$ and $P_m$ for
$\tau=2000$. Decaying dynamos are indicated by
black dots, \MD
dynamos are indicated by blue diamonds, \FD dynamos are indicated by
red stars, quadrupolar dynamos are indicated by green triangles, and
coexisting \MD and \FD dynamos are indicated by pink squares.
}
\label{fig01}
\end{figure*}

\section{\uppercase{Mathematical Formulation}}

We consider a spherical fluid shell rotating about a fixed vertical
axis. The basic equations permit a static solution with the
temperature distribution 
$$ 
T_S = T_0 + \Delta T \eta r^{-1} (1-\eta)^{-2},
$$
where $r$ denotes the distance from the center of the spherical shell
measured in multiples of the shell thickness $d$. The ratio
of inner to outer radius of the shell is denoted by $\eta$. $\Delta T$
is thus the temperature difference between the two boundaries.
The gravity field is given by 
$
\vec g = - d \gamma \vec r.
$
In addition to  $d$, the
time $d^2 / \nu$,  the temperature $\nu^2 / \gamma \alpha d^4$, and 
the magnetic flux density $\nu ( \mu \varrho )^{1/2} /d$ are used as
scales for the dimensionless description of the problem  where $\nu$ denotes
the kinematic viscosity of the fluid, $\kappa$ its thermal diffusivity,
$\varrho$ its density, and $\mu$ its magnetic permeability.
The equations of motion for the velocity vector $\vec u$, the heat
equation for the deviation 
$\Theta$ from the static temperature distribution, and the equation of
induction for the magnetic flux density $\vec B$ are thus given by 
\begin{gather}
\label{1a}
\partial_t \vec{u} + \vec u \cdot \nabla \vec u + \tau \vec k \times
\vec u = - \nabla \pi +\Theta \vec r + \nabla^2 \vec u + \vec B \cdot
\nabla \vec B, \\
\label{1b}
\nabla \cdot \vec u = 0, \\
\label{1c}
P(\partial_t \Theta + \vec u \cdot \nabla \Theta) = (R\, \eta
r^{-3} (1 - \eta)^{-2}) \vec r \cdot \vec u + \nabla^2 \Theta, \\
\nabla \cdot \vec B = 0, \\
\label{1d}
\nabla^2 \vec B =  P_m(\partial_t \vec B + \vec u \cdot \nabla \vec B
-  \vec B \cdot \nabla \vec u),
\end{gather}
where $\partial_t$ denotes the partial derivative with respect to time
$t$ and where all terms in the equation of motion that can be written
as gradients have been combined into $ \nabla \pi$. The Boussinesq
approximation is assumed in that the density $\varrho$ is
regarded as constant except in the gravity term where its temperature
dependence, given by $\alpha \equiv - ( \partial \varrho/\partial T)/\varrho =
$const., is taken into account. The Rayleigh numbers $R$,
the Coriolis number $\tau$, the Prandtl number $P$ and the magnetic
Prandtl number $P_m$ are defined by 
\begin{equation}
R = \frac{\alpha \gamma \Delta T d^4}{\nu \kappa},
\enspace \tau = \frac{2
\Omega d^2}{\nu} , \enspace P = \frac{\nu}{\kappa} , \enspace P_m = \frac{\nu}{\lambda},
\end{equation}
where $\lambda$ is the magnetic diffusivity.  Because the velocity 
field $\vec u$ as well as the magnetic flux density $\vec B$ are
solenoidal vector fields,   the general representation in terms of
poloidal and toroidal components can be used 
\begin{gather}
\vec u = \nabla \times ( \nabla v \times \vec r) + \nabla w \times 
\vec r \enspace , \\
\vec B = \nabla \times  ( \nabla h \times \vec r) + \nabla g \times 
\vec r \enspace .
\end{gather}
Equations for $v$ and $w$ are obtained by multiplication of the
curl$^2$ and of the curl of equation (1) by $\vec r$. Analogously
equations for $h$ and $g$ are obtained through the multiplication of
equation (5) and of its curl by $\vec r$. 

No-slip boundary conditions will be used at the inner boundary and
stress-free conditions are applied at the outer boundary, while the
temperature is assumed to be fixed at both boundaries, 
\begin{gather}
\label{ns}
v = \partial_r v = w = \Theta = 0
\quad \mbox{ at } r=r_i \equiv \eta/(1-\eta) 
\\ v = \partial^2_{rr}v = \partial_r (w/r) = \Theta =0
\quad \mbox{ at } r=r_o \equiv 1/(1-\eta).
\end{gather}
For the magnetic field, electrically insulating
boundaries are assumed such that the poloidal function $h$ must be 
matched to the function $h^{(e)}$, which describes the  
potential fields 
outside the fluid shell  
\begin{equation}
\label{mbc}
g = h-h^{(e)} = \partial_r ( h-h^{(e)})=0 \quad
\mbox{ at } r=r_i  \mbox{ and at}\quad r=r_o .
\end{equation}
Throughout this paper we shall use $\eta=0.65$.

The numerical integration proceeds with a pseudo-spectral 
method as described by Tilgner (1999) which is based on an expansion
of all dependent variables in spherical harmonics for the $\theta ,
\varphi$-dependences, i.e.  
\begin{equation}
v = \sum \limits_{l,m} V_l^m (r,t)\, P_l^m ( \cos \theta )\, \exp(im \varphi)
\end{equation}
and analogous expressions for $w, \Theta, h$ and $g$. 
Here $P_l^m$ denotes the associated Legendre functions.
For the $r$-dependence expansions in Chebychev polynomials are used. 
For the computations to be reported in the following a minimum of
41 collocation points in the radial direction and spherical harmonics
up to the order 128 have been used.

The numerical model outlined above is essentially the same as has been
used by \citet{GM81} and \citet{gi83}. At that time, 30 years ago,
computer resources allowed time integrations only for a few selected
parameter values and the phenomenon of bistability (see below) was not
yet known. It thus seems {appropriate} to return to the most
simple physically consistent formulation of convection driven dynamos
in rotating spherical fluid shells as formulated in this section. 

\section{\uppercase{Dynamo Solutions in Dependence on $P$ and} $P_m$}

In choosing parameter values in our model we follow \cite{gi83} in
selecting a value around $2000$ for $\tau$ which in the case of the
Sun corresponds to an eddy viscosity $\nu= 1.6\times
10^8$ m$^2$/s. Values for $R$ have usually been selected within a range
of up three times the value needed for dynamo action. Otherwise the dynamos
become too chaotic and regular features disappear. Our study has thus
focussed on the variations of dynamos with the parameters $P$ and
$P_m$.   

In figure 1 the existence of solutions for convection driven dynamos
in the $(P, P_m, R)$--space is outlined. In order to achieve dynamo
action the Rayleigh number $R$ must be sufficiently high such that the
magnetic Reynolds number $R_m$ exceeds a critical value. The latter is
inversely proportional to the magnetic Prandtl number $P_m$. Since our
study focusses on reasonably regular dynamos we have found it
difficult to obtain dynamos for small values of $P_m$ since dynamos
will become highly chaotic in this case or even decay. As indicated in
figure \ref{fig01} there exist a large variety of dynamos ranging from dipolar
to quadrupolar ones. At higher Rayleigh numbers the dynamos are often of mixed polarity.
While most of the dynamo solutions have been obtained for the radius
ratio $\eta=0.65$, some cases obtained for $\eta=0.6$ have been added
in the figure since it was found that the dynamos depend only weakly
on this parameter.   
\begin{figure}[t]
\vspace*{2mm}
\mypsfrag{Nui}{$\mathrm{Nu}_i$}
\mypsfrag{Mx}{$M_x$}
\mypsfrag{P}{$P$}
\begin{center}
\epsfig{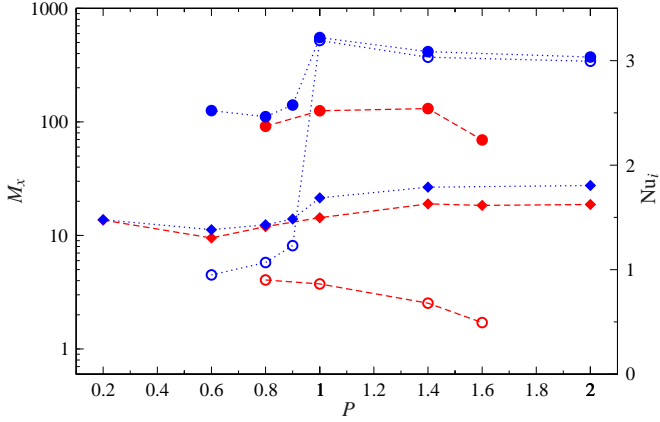}
\end{center}
\caption[]{(Color online)
Coexisting nonlinear dynamo solutions at identical parameter values
in the case $\eta=0.65$ $\tau=2000$, $R=150000$ and $P/P_m=0.2$.
The fluctuating poloidal magnetic energy component $\check{M}_p$
(solid circles) and the mean poloidal magnetic energy component
$\overline{M}_p$ (empty circles) are scaled on the left ordinate,
while the Nusselt number at $r=r_i$ (solid diamonds) is scaled on the
right ordinate. Cases started from initial conditions on the \FD branch
are shown in red {and connected by dashed lines}, and those started from
initial conditions on the \MD branch are shown in blue {and connected
by dotted lines.}} 
\label{fig02}
\end{figure}

\begin{figure}[ht]
\mypsfrag{Mmtp}{\scriptsize$\bar M_{t,p}$}
\mypsfrag{Mmp}{\scriptsize$\bar M_p$}
\mypsfrag{Mmt}{\scriptsize$\bar M_t$}
\mypsfrag{Mfp}{\scriptsize$\check M_p$}
\mypsfrag{Mft}{\scriptsize$\check M_t$}
\mypsfrag{Emp}{\scriptsize$\bar E_p$}
\mypsfrag{Emt}{\scriptsize$\bar E_t$}
\mypsfrag{Efp}{\scriptsize$\check E_p$}
\mypsfrag{Eft}{\scriptsize$\check E_t$}
\mypsfrag{Mdip}{\Mdip}
\mypsfrag{t}{$t$}
\mypsfrag{M}{$M$}
\mypsfrag{Ek}{$E$}
\mypsfrag{MD}{\MD}
\mypsfrag{FD}{\FD}
\mypsfrag{0}   {0}
\mypsfrag{0.2} {0.2}
\mypsfrag{0.4} {0.4}
\mypsfrag{0.6} {0.6}
\mypsfrag{0.8} {0.8}
\mypsfrag{1.2} {1.2}
\mypsfrag{1.6} {1.6}
\mypsfrag{1000}{1000}
\mypsfrag{2000}{2000}
\mypsfrag{3000}{3000}
\mypsfrag{4000}{4000}
\mypsfrag{5000}{5000}
\mypsfrag{6000}{6000}
\mypsfrag{10000}{10000}
\mypsfrag{(a)}{(a)}
\mypsfrag{(b)}{(b)}
\mypsfrag{(c)}{(c)}
\mypsfrag{(d)}{(d)}
\mypsfrag{(e)}{(e)}
\mypsfrag{(f)}{(f)}
\epsfig{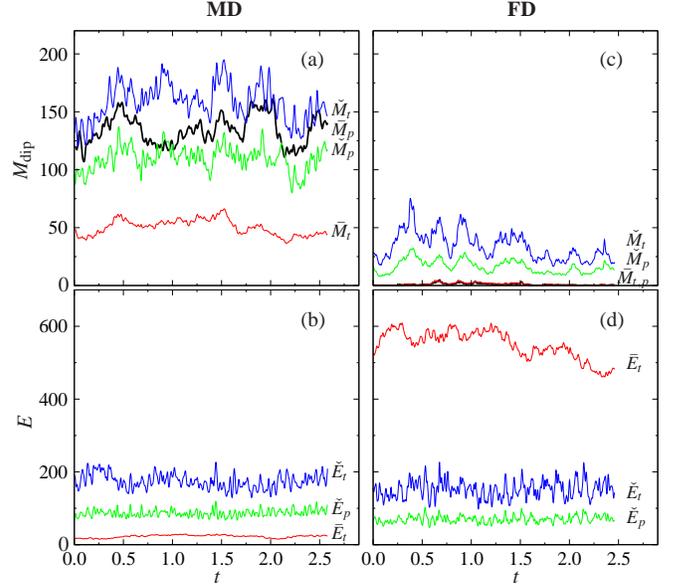}
\caption[]{(Color online)
Coexisting distinct {nonlinear dynamo solutions} at identical parameter
values -- a \MD (left column (a,b)) and a \FD dynamo (right column
(c,d)) both at $R=1.5\times10^6$, $\tau=2\times10^4$,
$P=1$ and  $P_m=5$.  Panels (a,c) and (b,d) show time series of
magnetic dipolar energy densities and kinetic energy densities,
respectively.
The component $\overline{X}_p$ is shown by thick solid black
line, while $\overline{X}_t$, $\check{X}_p$, and $\check{X}_t$
are shown by red, green and blue lines, respectively, and labeled. $X$ stands
for either $M$ or $E$.}
\label{fig03}
\end{figure}
Of particular interest is the distinction between \MD-dynamos which
are dominated by a strong and nearly steady axisymmetric dipolar
magnetic field and the \FD-dynamos for which magnetic energies of the
axisymmetric components are far less than those of the
non-axisymmetric components and which exhibit a cyclic behavior. Over
a considerable extent of the parameter range these two types of
dynamos exist simultaneously at the same values of the external
parameters as demonstrated by the example of  figure \ref{fig02}. It is remarkable that the two types of dynamos exhibit nearly the same Nusselt number. This property indicates that both types are similarly efficient in the use of the available buoyancy. For more details on the 
phenomenon of bistability we refer to the paper by 
\cite{sb09}. A comparison of the magnetic and kinetic energy densities
for the two types of dynamos is shown in figure \ref{fig03}. Here the
definitions  
\begin{eqnarray}
\label{7}
\bar M_p = \langle \mid \nabla \times ( \nabla \bar h \times \vec r )
\mid^2 \rangle /2 , \quad
& \bar M_t = \langle \mid \nabla \bar g \times \vec r \mid^2 \rangle/2
\\
\label{8}
\check M_p =  \langle \mid \nabla \times ( \nabla \check h
\times \vec r)
\mid^2 \rangle /2, \quad&
\check M_t = \langle \mid \nabla \check g \times\vec r \mid^2
\rangle/2
\end{eqnarray}
have been used where the angular brackets indicate the average over
the fluid shell  and where $\bar h$ refers to the azimuthally averaged 
component of $h$ and $\check h$ is given by $\check h = h -\bar h
$. Analogous relationships hold for the kinetic energy densities for
which $M, h, g$ are replaced by $E, v, w$.

Near the end points of the bistability regions transitions from \MD-
to \FD-dynamos and vice versa may, of course, occur owing to
exceptionally large fluctuations. Such a transition {was observed}
in the long-time dynamo integration of \cite{gd08}. As their convection
driven dynamo changed from the \MD-state to the \FD-state these
authors noted a solar like behavior in butterfly-like diagrams in
which they had plotted the axisymmetric $\varphi$-component of the
magnetic field, $\overline{B}_{\varphi}$, as a function of latitude and
time. As must be expected from the decrease with depth of their
differential rotation \citep{yo75}, the waves of the butterfly
diagram propagated to higher latitudes instead of lower ones. It turns
out, however, that non-axisymmetric components of the azimuthal
magnetic field are comparable in amplitude to the axisymmetric component
$\overline{B}_{\varphi}$. It thus appears to be appropriate to study
these dynamos in more detail. 
\begin{figure*}[ht]
\begin{center}
\epsfig{file=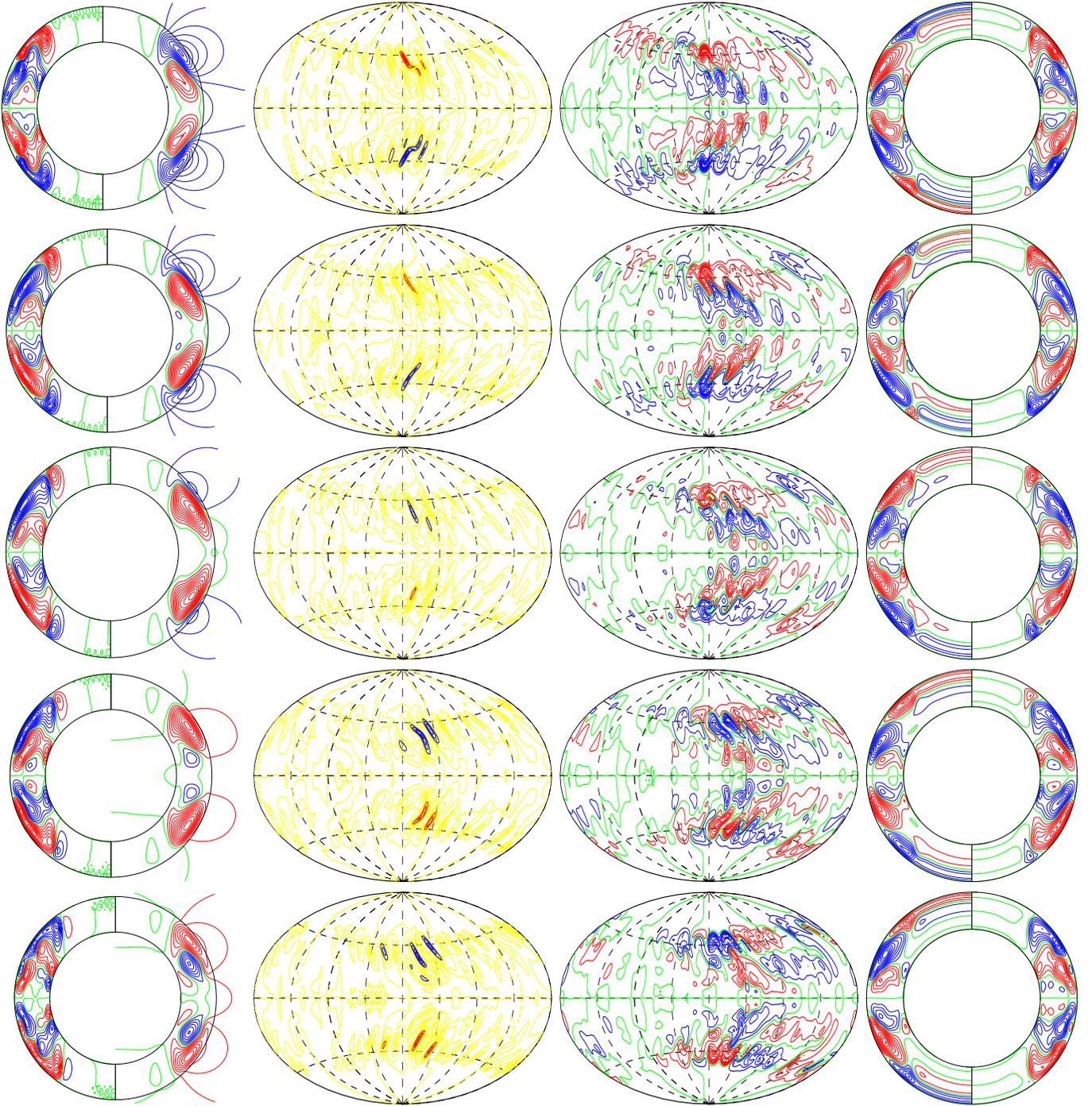,width=\textwidth,clip=}
\end{center}
\caption[]{{(Color online)} Approximately half a period of a dynamo oscillation in the case 
 $\tau=2000$, $R=120000$, $P=1.2$, $P_m=4.5$. The first
 column shows meridional lines of constant $\overline{B_{\varphi}}$ on
 the left and poloidal fieldlines, $r \sin\theta
 \partial\overline{h}/\partial\theta$ on the right. The second column shows
 lines of constant {horizontal magnetic field $(B_\varphi^2+B_\theta^2)^{1/2}$} at $r=0.9$
 corresponding to $-0.9$, $-0.8$, $-0.7$, $0.7$, $0.8$, $0.9$ of its maximum
 absolute value with the sign determined by the sign of {$B_{\varphi}$},  and the third column
 shows lines of constant $B_r$ at $r=r_o$. The last column shows
 {$\Re(\partial g^{m=1}/\partial\theta)$} to the left and
 {$\Im(\partial g^{m=1}/\partial\theta)$} to the right. The five
 rows are separated equidistantly in time by $\Delta t =0.0224$.} 
\label{fig04}
\end{figure*}

In figure \ref{fig04} a typical example of a nearly periodic \FD-dynamo is
shown. In the four columns of plots the development of various
components of the magnetic field is shown in five equidistant steps in
time covering about half a period of the magnetic cycle. As is evident
from the radial component $B_r$ displayed in the third column, the
dynamo appears to operate mainly in one of two meridional
hemispheres. In the second column maxima of the horizontal magnetic
field at the depth of 0.1 from the surface have been indicated as a
measure of the likelihood of the appearance of sunspots. Since the
underlying hemispherical structure hardly moves the pattern of the
second column seems to reflect the phenomenon of the active longitude
of sunspots \citetext{see, for example, \citet{u07} and references
therein} {or of solar X-ray flares \citep{z07}}. Since sometimes the $m=2$-component of the magnetic
field 
dominates instead of the $m=1$-component  the phenomenon of two active
longitudes 180 degrees apart can also be observed in the
dynamo simulations.   
Persistent active longitudes have been detected in various types of
active stars as well \citep{u07}. 

\begin{figure}
\epsfig{file=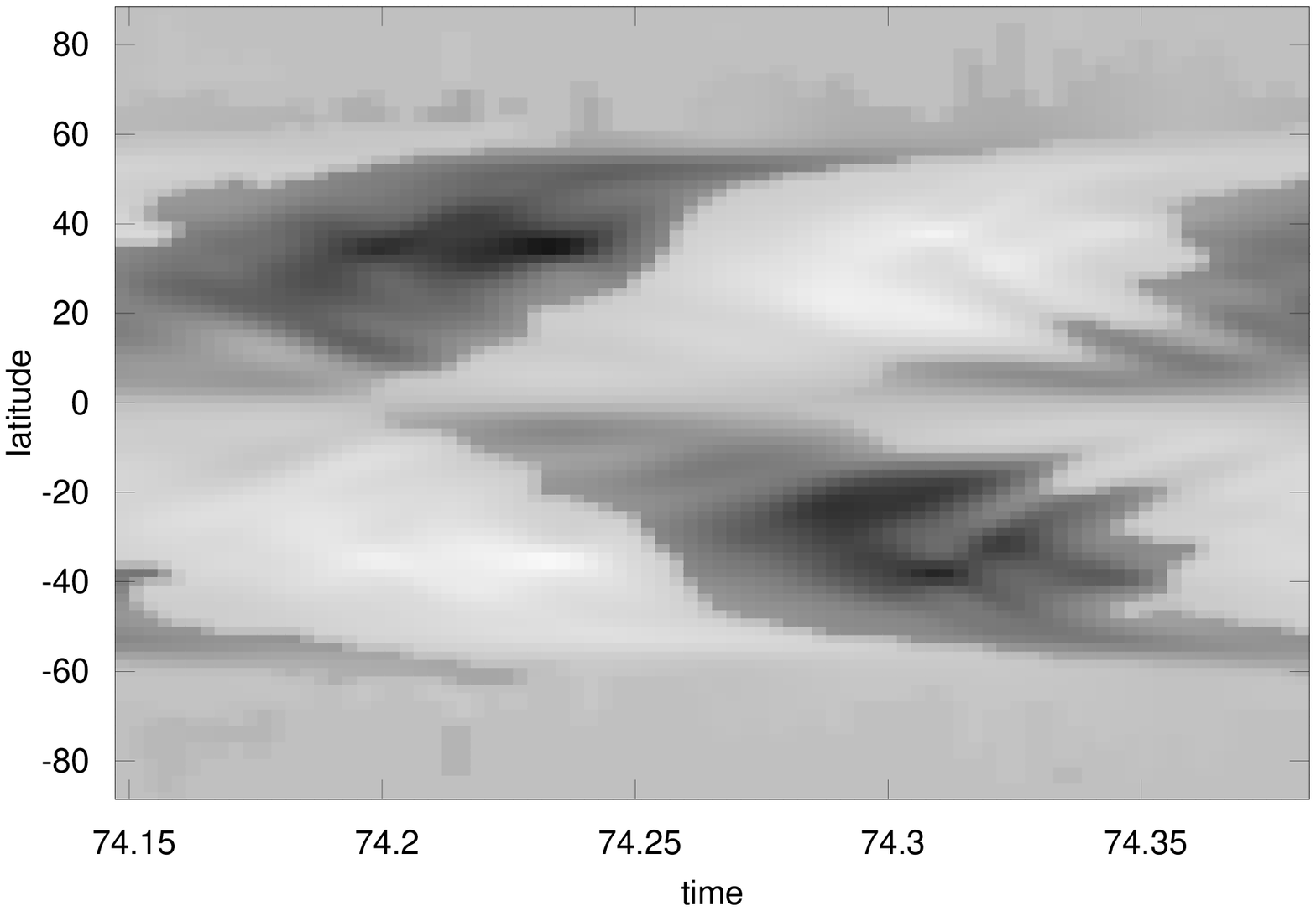,height=\columnwidth,angle=-90,clip=}\\
\epsfig{file=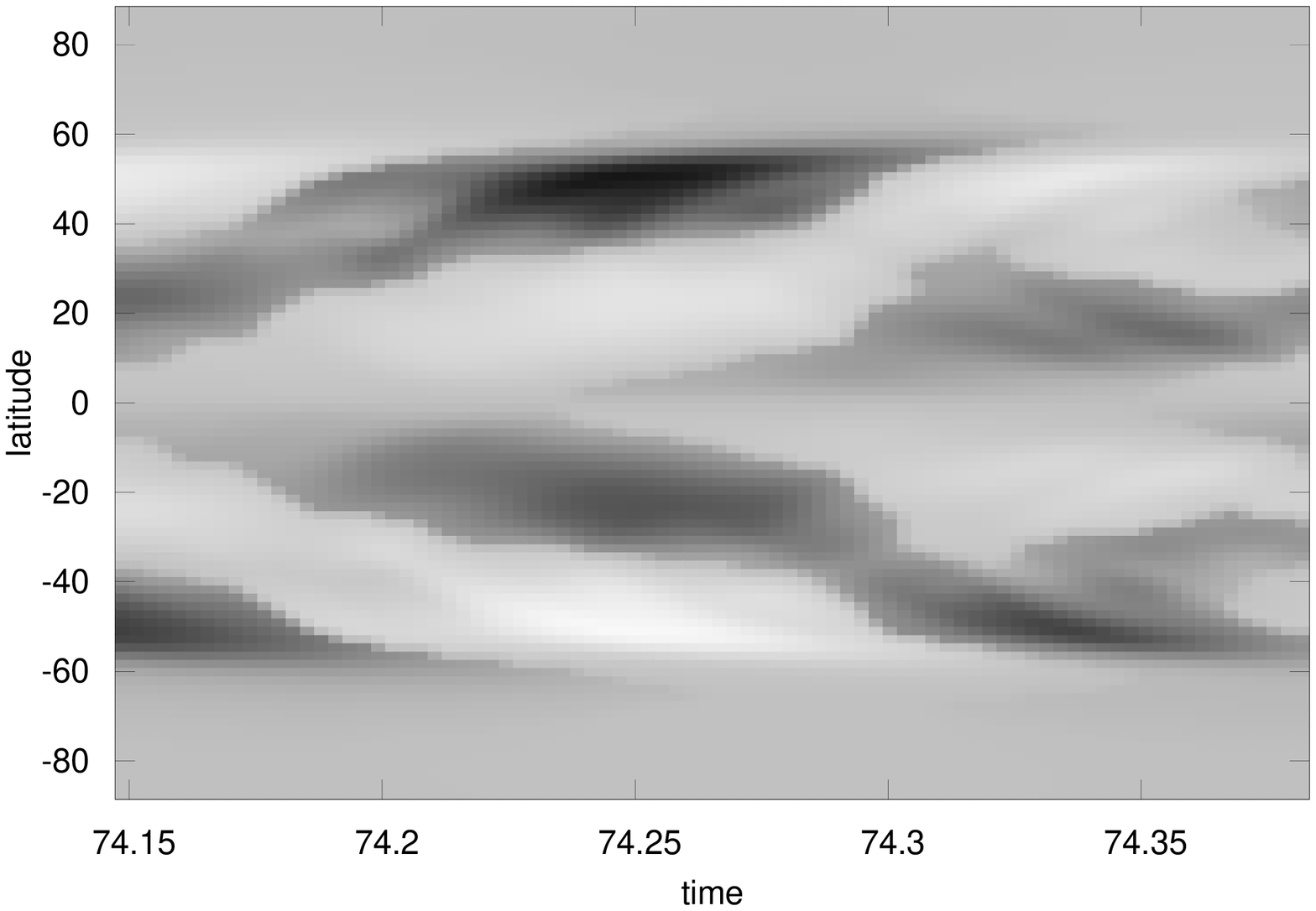,height=\columnwidth,angle=-90,clip=}\\
\caption[]{
"Butterfly" diagrams:
$B_\varphi^{m=0} + |B_\varphi^{m=1}| \sgn(B_\varphi^{m=0})$ (top) and
$B_r^{m=0} + |B_r^{m=1}| \sgn(B_r^{m=0})$ (bottom) 
are plotted as functions of latitude and time in the case $P=1.2$, $\tau=2000$, $R=120000$, $P_m=4.5$.
}
\label{fig05}
\end{figure}
In figure \ref{fig05} two "butterfly" diagrams have been
plotted. Because the strength of the axisymmetric component of
$B_{\varphi}$ alone is not representative of the horizontal magnetic
field that may give rise to sunspots we have added in the upper plot
the main non-axisymmetric component characterized by the azimuthal
wavenumber $m=1$. While there is some wave-like movement towards lower
latitudes, the main pattern moves towards high latitudes. This must be
expected since the differential rotation decreases with depth
\cite{yo75}. The movement toward higher latitudes is even more
pronounced in the pattern of the radial component $B_r$ of the
magnetic field as shown in the lower plot of figure \ref{fig05}. A phase shift
between the patterns of the upper and the lower plot can also be
noticed in that the sign change of $B_r$ occurs after that of
$B_{\varphi}$. This feature agrees with the well known solar
phenomenon that the  change in the sign of $B_r$ occurs near the
maximum of the sunspot cycle.     
        
\section{\uppercase{Conclusion}}

{Most models of the solar cycle are based on essentially axisymmetric
dynamos, but there are exceptions  such as the dynamo proposed by
\cite{r88} based on a mean field approach. It is not quite clear how
strong the non-axisymmetric components of the solar toroidal magnetic
field with wavenumbers $m=1$ or $m=2$ are in relation to the
axisymmetric component. Estimates range from $0.1$ based on sunspot
numbers \citep{u07} to $0.43$ based on powerful X-ray flares
\citep{z07}.    A main conclusion to be derived from {the} dynamo
model outlined in this 
paper is that the large scale non-axisymmetric components are likely to be an important ingredient of solar magnetism.  The
 well documented existence of active
longitudes certainly supports this conclusion.
}

There are, of course, also major differences between solar phenomena
and certain features of the present Boussinesq model. Some of these are well understood as
for example the difference in the propagation of dynamo waves. This is
most probably caused by the fact that the model always exhibits a
differential rotation decreasing with depth for \FD-type dynamos,
while the Sun possesses a differential rotation that increases with
depth in the uppermost $15\%$ of the convection zone. For a
preliminary attempt to take this effect into account we refer to a
recent companion paper by \cite{sb12}. 
  
\acknowledgments
The research of R.S.~has been partly supported by the UK Royal 
Society under Research Grant 2010 R2. The research of F.B.~has been
supported by NASA Grant NNX09AJ85G.



\end{document}